# Phylotrack: C++ and Python libraries for *in silico* phylogenetic tracking

**Emily Dolson** [1,2,¶], **Santiago Rodruigez-Papa** [1], **and Matthew Andres Moreno** [3,4,5]

1 Department of Computer Science and Engineering, Michigan State University, East Lansing, MI, 48823, USA  2 Ecology, Evolution, and Behavior, Michigan State University, East Lansing, MI, 48823, USA  3 Department of Ecology and Evolutionary Biology, University of Michigan, Ann Arbor, MI, 48103 USA  4 Center for the Study of Complex Systems, University of Michigan, Ann Arbor, MI, 48103, USA  5 Michigan Institute for Data Science and AI in Society, University of Michigan, Ann Arbor, MI, 48103, USA ¶ Corresponding author

## Abstract

### Summary

Phylotrack provides efficient C++ and Python libraries for tracking and analyzing phylogenies generated *in silico*. It offers flexible configuration of operational taxonomic units, memory-saving pruning of extinct lineages, and calculation of common phylogeny metrics.

### Availability and Implementation

Source code is freely available on GitHub (https://github.com/emilydolson/phylotrackpy) and archived via Zenodo (https://doi.org/10.5281/zenodo.10888780). The *phylotrackpy* Python package can be installed via PyPI (https://pypi.org/project/phylotrackpy/). Documentation hosted on ReadTheDocs (https://phylotrackpy.readthedocs.io/). Phylotrack is provided under the MIT License.

### Contact

Emily Dolson, dolsonem@msu.edu

### Supplementary Information

Profiling data are available via the Open Science Framework at https://osf.io/52hzs/.

## Introduction

*In silico* evolution instantiates the processes of heredity, variation, and differential reproductive success (the three "ingredients" for evolution by natural selection) within digital populations of computational agents. Consequently, these populations undergo evolution (Pennock, 2007), and can be used as virtual model systems for studying evolutionary dynamics. This experimental paradigm — used across biological modeling, artificial life, and evolutionary computation — complements research done using *in vitro* and *in vivo* systems by enabling experiments that would be impossible in the lab or field (E. Dolson & Ofria, 2021). Although phylogenetic data from *in vivo* systems is growing ever more fine-grained (Konno et al., 2022), complete, exact observability remains a key benefit of computational methods. Indeed, in digital populations it is possible to perfectly record all parent-child relationships across history, yielding complete phylogenies (ancestry trees). This information reveals when traits were gained or lost, and also facilitates inference of underlying evolutionary dynamics (E. Dolson et al., 2020; Lewinsohn et al., 2023; Mooers & Heard, 1997; Moreno, Rodriguez-Papa, et al., 2024; Scott et al., 2018).

Such capability is of great value to phylogenetics research. While backward-time models such as coalescent-based approaches are powerful, there are many scenarios that they cannot support, necessitating the use of forward-time evolutionary models (Benjamin C. Haller & Messer, 2019). Thus, research agendas that require predicting the phylogenetic patterns that would result from a given evolutionary process would benefit greatly from a way to track phylogenies in forward-time *in silico* evolution. Currently, such tracking is implemented case-by-case in individual digital model systems. As efficient phylogenetic tracking is non-trivial to implement, this *status quo* burdens research workloads.

To solve this problem, we have developed the Phylotrack project, which provides libraries for tracking and analyzing phylogenies in *in silico* evolution. The project is composed of 1) Phylotracklib: a header-only C++ library, developed under the umbrella of the Empirical project (Ofria et al., 2020), and 2) Phylotrackpy: a Python wrapper around Phylotracklib, created with Pybind11 (Jakob et al., 2017). Both components supply a public-facing API to attach phylogenetic tracking to digital evolution systems, as well as a stand-alone interface for measuring a variety of popular phylogenetic topology metrics (Tucker et al., 2017). Underlying design and C++ implementation prioritize efficiency, allowing for fast generational turnover for agent populations numbering in the tens of thousands. Several explicit features (e.g., phylogeny pruning and abstraction, etc.) are provided to reduce the memory footprint of phylogenetic information.

**Related work**

*In silico* evolution work enjoys a rich history of phylogenetic measurement and analysis, and many systems facilitate tracking phylogenies (Bohm et al., 2017; De Rainville et al., 2012; Garwood et al., 2019; Ofria & Wilke, 2004; Ray, 1992). In particular, the bioinformatics community has fielded a rich set of agent-based, forward-time simulation frameworks capable of reporting evolutionary history. These include Nemo (Guillaume & Rougemont, 2006), TreeSimJ (O'Fallon, 2010), fwdpp (Thornton, 2014), sPEGG (Okamoto & Amarasekare, 2017), MimicrEE2 (Vlachos & Kofler, 2018), hexsim (Schumaker & Brookes, 2018), SimBit (Matthey-Doret, 2021), and SLiM (Benjamin C. Haller & Messer, 2023). SLiM is a popular — and powerful — exemplar of the framework-oriented approach. Simulations are highly configurable, owing to inclusion of an inbuilt scripting language (B. Haller, 2016), yet still ultimately must take place within SLiM's tick cycle and community-species-subpopulation ontology. Such an approach enables highly optimized operation and reduces end-user workload. In contrast, Phylotrack provides ready-built tracking flexible enough to attach to any population of digital replicating entities, however implemented or organized. To our knowledge, no other general-purpose phylogeny tracking library currently exists, as prior work has relied on bespoke system- or framework-specific implementations.

Two other general-purpose libraries for phylogenetic record-keeping do exist: hstrat and Automated Phylogeny Over Geological Timescales (APOGeT). However, they provide different modes of phylogenetic instrumentation than Phylotrack does. Whereas Phylotrack uses a graph-based approach to perfectly record asexual phylogenies, the hstrat library implements hereditary stratigraphy, a technique for decentralized phylogenetic tracking that is approximate instead of exact (Moreno et al., 2022b) (see (Moreno, Papa, et al., 2024) for a more thorough comparison). APOGeT, in turn, focuses on tracking speciation in sexually reproducing populations (Godin-Dubois et al., 2019).

Vast amounts of bioinformatics-oriented phylogenetic analysis software is also available. Applications typically include

- inferring phylogenies from extant organisms (and sometimes fossils) (Challa & Neelapu,

2019),
- sampling phylogenies from theoretical models of population and species dynamics (Stadler, 2011),
- cross-referencing phylogenies with other data (e.g., spatial species distributions) (Emerson & Gillespie, 2008), and
- analyzing and manipulating tree structures (Cock et al., 2009; Moreno, Holder, et al., 2024; Sand et al., 2014; Smith, 2020).

Phylotrack overlaps with these goals only in that it also provides tree statistic implementations. We chose to include this feature to facilitate fast during-simulation calculations of these metrics. Notably, the problem of tracking a phylogeny within an agent-based program differs substantially from the more traditional problem of reconstructing a phylogeny. Users new to recorded phylogenies should refer to the Phylotrackpy documentation[1] for notes on subtle structural differences from reconstructed phylogenies.

Phylotrack has contributed to a variety of published research projects through integrations with Modular Agent-Based Evolver (MABE) 2.0 (Bohm et al., 2019), Symbulation (Vostinar & Ofria, 2019), and even a fork of the Avida digital evolution platform (E. Dolson et al., 2020; Ofria & Wilke, 2004). Research topics include open-ended evolution (E. L. Dolson et al., 2019), the origin of endosymbiosis (Johnson et al., 2022), the importance of phylogenetic diversity for machine learning via evolutionary computation (Hernandez et al., 2022; Shahbandegan et al., 2022), and more. Phylotrackpy is newer, but it has already served as a point of comparison in the development of other phylogenetic tools (Moreno et al., 2022a, 2023).

---

[1]Phlotrack documentation is hosted via ReadTheDocs at https://phylotrackpy.readthedocs.io/.

# Methods

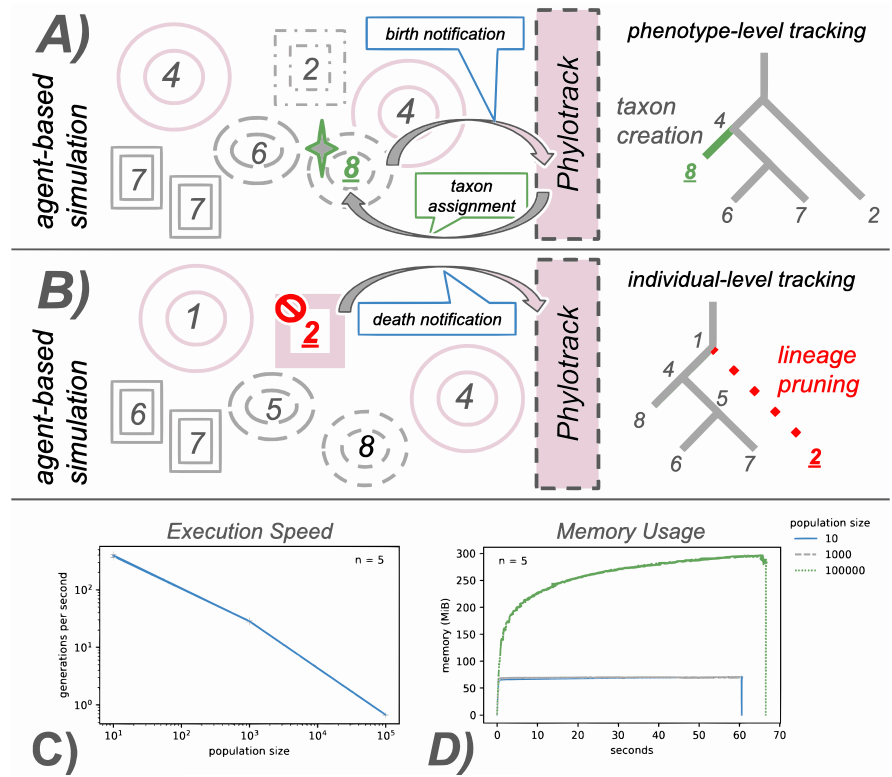

**Figure 1:** Phylotrack lineage recording operations and performance profiling results. Panel A overviews taxon creation, in which end-user simulation code reports an agent birth to Phylotrack. Panel B overviews taxon removal, in which end-user simulation code reports an agent death to Phylotrack. Note that panels A and B differ in configured operational taxonomic unit. In the first, individual agents define the taxonomic unit and, in the second, agent phenotype classes define the taxonomic unit. Panel C shows execution speed across population sizes. Panel D shows allocated memory over a 60-second execution window. Error bars are SE.

## Lineage Recording

The core functionality of Phylotrack is to record asexual phylogenies from simulation agent creation and destruction events. To improve memory efficiency, extinct branches are pruned from phylogenies by default, but this feature can be disabled. Further efficiency can be gained by coarsening the level of abstraction applied (i.e., what constitutes a taxonomic unit). Configurability of the operational taxonomic unit, in addition, helps users collect data that matches their experimental objectives. For instance, recent work on phylometric signals of evolutionary conditions found substantial qualitative differences in phylometrics collected from individual- versus genotype-level tracking ([Moreno, Rodriguez-Papa, et al., 2024](#)). Supplemental data about each taxonomic unit can be stored efficiently.

Top panels of Figure [1](#) overview Phylotrack's two core operations: taxon creation and removal. On simulation startup, end-users report founding agents to Phylotrack. Each is assigned a Taxon ID. Subsequently, as shown in Figure [1](#)A, user code informs Phylotrack of simulation reproduction events as they occur. Given (1) the parent's Taxon ID and (2) the offspring's taxonomic traits (e.g., genome for genotype-level tracking, trait vector for phenotype-level tracking, etc.), Phylotrack returns a Taxon ID for the offspring. If the offspring's taxonomic traits match its parent, their Taxon ID will be identical; otherwise, Phylotrack assigns the offspring a new Taxon ID. Figure [1](#)B shows the other primary Phylotrack operation, taxon removal. Each time simulation discards an agent, user code reports its Taxon ID to Phylotrack. If no extant agents carry that Taxon ID and the Taxon ID has no extant descendants, Phylotrack conducts a pruning operation, deleting corresponding lineage records to reclaim memory (unless the user has disabled pruning).

Lineage recording in Phylotrack is efficient. The worst-case time complexity of taxon creation is O(1) with respect to both population size and generations elapsed ([Moreno, Papa, et al., 2024](#)). In pruning-enabled tracking, taxon removal is amortized O(1). Space complexity is harder to calculate meaningfully, but should be O(N) on average in most evolutionary scenarios (where N is population size). For details, refer to Moreno, Papa, et al. ([2024](#)).

### Serialization

Phylotrack exports data in the Artificial Life Standard Phylogeny format ([Lalejini et al., 2019](#)). This format integrates with an associated ecosystem of software converters, analyzers, and visualizers. Existing tools support easy conversion to bioinformatics-standard formats (e.g., Newick, phyloXML, etc.) ([Papa, 2024](#)), allowing Phylotrack phylogenies to be analyzed with tools designed for biological data. Phylogeny data can be restored from file, enabling post-hoc calculation of phylogenetic topology statistics.

### Phylogenetic Topology Statistics

Support is provided for

- Average phylogenetic depth across taxa
- Average origin time across taxa
- Most recent common ancestor origin time
- Shannon diversity ([Spellerberg & Fedor, 2003](#))
- Colless-like index ([Mir et al., 2018](#))
- Mean, sum, and variance of evolutionary distinctiveness ([Isaac et al., 2007](#); [Tucker et al., 2017](#))
- Mean, sum, and variance pairwise distance ([Clarke & Warwick, 1998](#), [2001](#); [Tucker et al., 2017](#); [Webb et al., 2002](#))
- Phylogenetic diversity ([Faith, 1992](#))
- Sackin's index ([Shao & Sokal, 1990](#))

### Profiling

To assess Phylotrack performance, we ran a simple asexual evolutionary algorithm instrumented with lineage-pruned systematics tracking. We performed neutral selection with a 20% mutation probability. Genomes consisted of a single floating-point value, which also served as the taxonomic unit.

We ran 60-second trials using population sizes 10, 1,000, and 100,000, with five replicates each. Each trial concluded with a data export operation. We used `memory_profiler` (`psutil` backend) to measure process memory usage ([Pedregosa & Gervais, 2023](#)). Full profiling data and hardware specifications hosted via the Open Science Framework at [https://osf.io/52hzs/](https://osf.io/52hzs/) ([Foster & Deardorff, 2017](#)).

**Execution Speed**

Figure [1](#)C shows generations evaluated per second at each population size. At population size 10, 1,000, and 100,000, we observed 3,923 (s.d. 257), 28,386 (s.d. 741), and 67,000 (s.d. 1,825) agent reproduction events per second. Efficiency gains with population size likely arose from NumPy vectorized operations used to perform mutation and selection.

**Memory Usage**

Phylotrack consumes 296 MiB (s.d. 1.1) peak memory to track a population of 100,000 agents over 40 (s.d. 1) generations. At population sizes 10 and 1,000, peak memory usage was 70.6 MiB (s.d. 0.5) and 71.0 MiB (s.d. 0.2). Figure [1]D shows memory use trajectories over 60-second trials.

Most applications should expect lower memory usage because selection typically increases opportunities for lineage pruning.

**Availability and Implementation**

Phylotrackpy's source code is open source at https://github.com/emilydolson/phylotrackpy. It is provided under the MIT License. Installation — including cross-platform precompiled wheel files — can be performed via PyPI.[2] To ensure perpetual availability, PhylotrackPy and its C++ backend are archived via Zenodo (E. Dolson et al., 2024; Ofria et al., 2020).

## Discussion

Agent-based, forward-time modeling has an important role to play in integrating and extending our understanding of natural systems (Marshall & Galea, 2014; Urban et al., 2021). Phylotrack seeks to complement the rich collection of existing framework-oriented agent-based simulation systems by providing a library tool that empowers Python and C++ users to extract evolutionary history from fully custom projects. Such an approach, in particular, supports end-user work that also leverages other tools from the Python package ecosystem (Perez et al., 2011; Raschka et al., 2020) and reaches a broad, diverse audience of Python users (Perkel, 2015). Our ultimate goal in releasing this software is to better bring bioinformatics to bear in digital evolution experiments, and to better recruit digital evolution in generating synthetic data that supports bioinformatics work (Sukumaran et al., 2021).

In future work, we plan to extend Phylotrack in a future release to allow multiple parents per taxon. A current limitation of Phylotrack is incompatibility with sexually reproducing populations (unless tracking is per gene). We also look forward to expanding Phylotrack's portfolio of natively-implemented phylogeny structure metrics.

## Acknowledgements

This research was supported by Michigan State University through the computational resources provided by the Institute for Cyber-Enabled Research; and the Eric and Wendy Schmidt AI in Science Postdoctoral Fellowship, a Schmidt Futures program [to MAM].

## References

Bohm, C., G., N. C., & Hintze, A. (2017). *MABE (modular agent based evolver): A framework for digital evolution research. 14*, 76–83. https://doi.org/10.7551/ecal_a_016

Bohm, C., Lalejini, A., Schossau, J., & Ofria, C. (2019). MABE 2.0: An introduction to MABE and a road map for the future of MABE development. *Proceedings of the Genetic and Evolutionary Computation Conference Companion*, 1349–1356. https://doi.org/10.1145/3319619.3326825

Challa, S., & Neelapu, N. R. R. (2019). Phylogenetic trees: Applications, construction, and assessment. *Essentials of Bioinformatics, Volume III: In Silico Life Sciences: Agriculture*, 167–192. https://doi.org/10.1007/978-3-030-19318-8_10

---

[2]Phylotrack's PyPI package listing is at https://pypi.org/project/phylotrackpy/.

Clarke, K. R., & Warwick, R. M. (1998). Quantifying structural redundancy in ecological communities. *Oecologia*, *113*(2), 278–289. https://doi.org/10.1007/s004420050379

Clarke, K. R., & Warwick, R. M. (2001). A further biodiversity index applicable to species lists: Variation in taxonomic distinctness. *Marine Ecology Progress Series*, *216*, 265–278. https://doi.org/10.3354/meps216265

Cock, P. J., Antao, T., Chang, J. T., Chapman, B. A., Cox, C. J., Dalke, A., Friedberg, I., Hamelryck, T., Kauff, F., Wilczynski, B., & others. (2009). Biopython: Freely available python tools for computational molecular biology and bioinformatics. *Bioinformatics*, *25*(11), 1422–1423. https://doi.org/10.1093/bioinformatics/btp163

De Rainville, F.-M., Fortin, F.-A., Gardner, M.-A., Parizeau, M., & Gagné, C. (2012). DEAP: A Python framework for evolutionary algorithms. *Proceedings of the 14th Annual Conference Companion on Genetic and Evolutionary Computation*, 85–92. https://doi.org/10.1145/2330784.2330799

Dolson, E. L., Vostinar, A. E., Wiser, M. J., & Ofria, C. (2019). The MODES toolbox: Measurements of open-ended dynamics in evolving systems. *Artificial Life*, *25*(1), 50–73. https://doi.org/10.1162/artl_a_00280

Dolson, E., Lalejini, A., Jorgensen, S., & Ofria, C. (2020). Interpreting the tape of life: Ancestry-based analyses provide insights and intuition about evolutionary dynamics. *Artificial Life*, *26*(1), 1–22. https://doi.org/10.1162/artl_a_00313

Dolson, E., Moreno, M. A., & rodsan0. (2024). *Emilydolson/phylotrackpy: v0.2.0* (Version v0.2.0). Zenodo. https://doi.org/10.5281/zenodo.10888780

Dolson, E., & Ofria, C. (2021). Digital evolution for ecology research: A review. *Frontiers in Ecology and Evolution*, *9*, 852. https://doi.org/10.3389/fevo.2021.750779

Emerson, B. C., & Gillespie, R. G. (2008). Phylogenetic analysis of community assembly and structure over space and time. *Trends in Ecology & Evolution*, *23*(11), 619–630. https://doi.org/10.1016/j.tree.2008.07.005

Faith, D. P. (1992). Conservation evaluation and phylogenetic diversity. *Biological Conservation*, *61*(1), 1–10. https://doi.org/10.1016/0006-3207(92)91201-3

Foster, E. D., & Deardorff, A. (2017). Open science framework (OSF). *Journal of the Medical Library Association: JMLA*, *105*(2), 203. https://doi.org/10.5195/jmla.2017.88

Garwood, R. J., Spencer, A. R. T., & Sutton, M. D. (2019). REvoSim: Organism-level simulation of macro- and microevolution. *Palaeontology*, *62*(3), 339–355. https://doi.org/10.1111/pala.12420

Godin-Dubois, K., Cussat-Blanc, S., & Duthen, Y. (2019, August). *APOGeT: Automated phylogeny over geological timescales*. https://doi.org/10.13140/rg.2.2.33781.93921

Guillaume, F., & Rougemont, J. (2006). Nemo: An evolutionary and population genetics programming framework. *Bioinformatics*, *22*(20), 2556–2557. https://doi.org/10.1093/bioinformatics/btl415

Haller, B. (2016). Eidos: A simple scripting language. *URL: Http://Benhaller.com/Slim/Eidos_Manual.pdf*.

Haller, Benjamin C., & Messer, P. W. (2019). SLiM 3: Forward Genetic Simulations Beyond the Wright–Fisher Model. *Molecular Biology and Evolution*, *36*(3), 632–637. https://doi.org/10.1093/molbev/msy228

Haller, Benjamin C., & Messer, P. W. (2023). SLiM 4: Multispecies eco-evolutionary modeling. *The American Naturalist*, *201*(5), E127–E139. https://doi.org/10.1086/723601

Hernandez, J. G., Lalejini, A., & Dolson, E. (2022). What can phylogenetic metrics tell us

about useful diversity in evolutionary algorithms? In W. Banzhaf, L. Trujillo, S. Winkler, & B. Worzel (Eds.), *Genetic programming theory and practice XVIII* (pp. 63–82). Springer Nature Singapore. https://doi.org/10.1007/978-981-16-8113-4_4

Isaac, N. J. B., Turvey, S. T., Collen, B., Waterman, C., & Baillie, J. E. M. (2007). Mammals on the EDGE: Conservation priorities based on threat and phylogeny. *Plos One*, *2*(3), e296. https://doi.org/10.1371/journal.pone.0000296

Jakob, W., Rhinelander, J., & Moldovan, D. (2017). *pybind11 – seamless operability between C++11 and Python*. https://github.com/pybind/pybind11

Johnson, K., Welch, P., Dolson, E., & Vostinar, A. E. (2022, July 18). *Endosymbiosis or bust: Influence of ectosymbiosis on evolution of obligate endosymbiosis*. ALIFE 2022: The 2022 conference on artificial life. https://doi.org/10.1162/isal_a_00488

Konno, N., Kijima, Y., Watano, K., Ishiguro, S., Ono, K., Tanaka, M., Mori, H., Masuyama, N., Pratt, D., Ideker, T., Iwasaki, W., & Yachie, N. (2022). Deep distributed computing to reconstruct extremely large lineage trees. *Nature Biotechnology*, *40*(4), 566–575. https://doi.org/10.1038/s41587-021-01111-2

Lalejini, A., Dolson, E., Bohm, C., Ferguson, A. J., Parsons, D. P., Rainford, P. F., Richmond, P., & Ofria, C. (2019). Data standards for artificial life software. *ALIFE 2019: The 2019 Conference on Artificial Life*, 507–514. https://doi.org/10.1162/isal_a_00213

Lewinsohn, M. A., Bedford, T., Müller, N. F., & Feder, A. F. (2023). State-dependent evolutionary models reveal modes of solid tumour growth. *Nature Ecology & Evolution*, *7*(4), 581–596. https://doi.org/10.1038/s41559-023-02000-4

Marshall, B. D. L., & Galea, S. (2014). Formalizing the role of agent-based modeling in causal inference and epidemiology. *American Journal of Epidemiology*, *181*(2), 92–99. https://doi.org/10.1093/aje/kwu274

Matthey-Doret, R. (2021). SimBit: A high performance, flexible and easy-to-use population genetic simulator. *Molecular Ecology Resources*, *21*(5), 1745–1754. https://doi.org/10.1111/1755-0998.13372

Mir, A., Rotger, L., & Rosselló, F. (2018). Sound colless-like balance indices for multifurcating trees. *Plos One*, *13*(9), e0203401. https://doi.org/10.1371/journal.pone.0203401

Mooers, A. O., & Heard, S. B. (1997). Inferring evolutionary process from phylogenetic tree shape. *The Quarterly Review of Biology*, *72*(1), 31–54. https://doi.org/10.1086/419657

Moreno, M. A., Dolson, E., & Ofria, C. (2022a). Hereditary stratigraphy: Genome annotations to enable phylogenetic inference over distributed populations. *ALIFE 2022: The 2022 Conference on Artificial Life*, 65–66. https://doi.org/10.1162/isal_a_00550

Moreno, M. A., Dolson, E., & Ofria, C. (2022b). Hstrat: A python package for phylogenetic inference on distributed digital evolution populations. *Journal of Open Source Software*, *7*(80), 4866. https://doi.org/10.21105/joss.04866

Moreno, M. A., Dolson, E., & Rodriguez-Papa, S. (2023). *Toward Phylogenetic Inference of Evolutionary Dynamics at Scale*. ALIFE 2023: Ghost in the Machine: Proceedings of the 2023 Artificial Life Conference, 79. https://doi.org/10.1162/isal_a_00694

Moreno, M. A., Holder, M. T., & Sukumaran, J. (2024). *DendroPy 5: A mature python library for phylogenetic computing*. https://doi.org/10.48550/arXiv.2405.14120

Moreno, M. A., Papa, S. R., & Dolson, E. (2024). *Analysis of phylogeny tracking algorithms for serial and multiprocess applications*. arXiv. https://doi.org/10.48550/arXiv.2403.00246

Moreno, M. A., Rodriguez-Papa, S., & Dolson, E. (2024). *Ecology, spatial structure, and selection pressure induce strong signatures in phylogenetic structure*. https://doi.org/10.48550/arXiv.2405.07245

O'Fallon, B. (2010). TreesimJ: A flexible, forward time population genetic simulator. *Bioinformatics*, *26*(17), 2200–2201. https://doi.org/10.1093/bioinformatics/btq355

Ofria, C., Moreno, M. A., Dolson, E., Lalejini, A., Rodriguez Papa, S., Fenton, J., Perry, K., Jorgensen, S., hoffmanriley, grenewode, Baldwin Edwards, O., Stredwick, J., cgnitash, theycallmeHeem, Vostinar, A., Moreno, R., Schossau, J., Zaman, L., & djrain. (2020). *Empirical: C++ library for efficient, reliable, and accessible scientific software* (Version 0.0.4). https://doi.org/10.5281/zenodo.4141943

Ofria, C., & Wilke, C. O. (2004). Avida: A software platform for research in computational evolutionary biology. *Artificial Life*, *10*(2), 191–229. https://doi.org/10.1162/106454604773563612

Okamoto, K. W., & Amarasekare, P. (2017). A framework for high-throughput eco-evolutionary simulations integrating multilocus forward-time population genetics and community ecology. *Methods in Ecology and Evolution*, *9*(3), 525–534. https://doi.org/10.1111/2041-210x.12889

Papa, M. A. M. A. S. R. (2024). *mmore500/alifedata-phyloinformatics-convert*. Zenodo. https://doi.org/10.5281/zenodo.10701178

Pedregosa, F., & Gervais, P. (2023). *Memory profiler*.

Pennock, R. T. (2007). Models, simulations, instantiations, and evidence: The case of digital evolution. *Journal of Experimental & Theoretical Artificial Intelligence*, *19*(1), 29–42. https://doi.org/10.1080/09528130601116113

Perez, F., Granger, B. E., & Hunter, J. D. (2011). Python: An ecosystem for scientific computing. *Computing in Science &Amp; Engineering*, *13*(2), 13–21. https://doi.org/10.1109/mcse.2010.119

Perkel, J. M. (2015). Programming: Pick up python. *Nature*, *518*(7537), 125–126. https://doi.org/10.1038/518125a

Raschka, S., Patterson, J., & Nolet, C. (2020). Machine learning in python: Main developments and technology trends in data science, machine learning, and artificial intelligence. *Information*, *11*(4), 193. https://doi.org/10.3390/info11040193

Ray, T. (1992). Evolution, ecology and optimization of digital organisms. *Santa Fe Institute Working Paper*, *92*. https://homeostasis.scs.carleton.ca/~soma/adapsec/readings/tierra-92-08-042.pdf

Sand, A., Holt, M. K., Johansen, J., Brodal, G. S., Mailund, T., & Pedersen, C. N. (2014). tqDist: A library for computing the quartet and triplet distances between binary or general trees. *Bioinformatics*, *30*(14), 2079–2080. https://doi.org/10.1093/bioinformatics/btu157

Schumaker, N. H., & Brookes, A. (2018). HexSim: A modeling environment for ecology and conservation. *Landscape Ecology*, *33*(2), 197–211. https://doi.org/10.1007/s10980-017-0605-9

Scott, J. G., Maini, P. K., Anderson, A. R. A., & Fletcher, A. G. (2018). *Inferring tumour proliferative organisation from phylogenetic tree measures in a computational model*. https://doi.org/10.1101/334946

Shahbandegan, S., Hernandez, J. G., Lalejini, A., & Dolson, E. (2022). Untangling phylogenetic diversity's role in evolutionary computation using a suite of diagnostic fitness landscapes. *Proceedings of the Genetic and Evolutionary Computation Conference Companion*, 2322–2325. https://doi.org/10.1145/3520304.3534028

Shao, K.-T., & Sokal, R. R. (1990). Tree balance. *Systematic Zoology*, *39*(3), 266–276. https://doi.org/10.2307/2992186

Smith, M. R. (2020). TreeDist: Distances between phylogenetic trees. R package version

2.6.1. In *Comprehensive R Archive Network*. https://doi.org/10.5281/zenodo.3528124

Spellerberg, I. F., & Fedor, P. J. (2003). A tribute to claude shannon (1916–2001) and a plea for more rigorous use of species richness, species diversity and the 'shannon–wiener' index. *Global Ecology and Biogeography*, *12*(3), 177–179. https://doi.org/10.1046/j.1466-822X.2003.00015.x

Stadler, T. (2011). Simulating trees with a fixed number of extant species. *Systematic Biology*, *60*(5), 676–684. https://doi.org/10.1093/sysbio/syr029

Sukumaran, J., Holder, M. T., & Knowles, L. L. (2021). Incorporating the speciation process into species delimitation. *PLOS Computational Biology*, *17*(5), e1008924. https://doi.org/10.1371/journal.pcbi.1008924

Thornton, K. R. (2014). A c++ template library for efficient forward-time population genetic simulation of large populations. *Genetics*, *198*(1), 157–166. https://doi.org/10.1534/genetics.114.165019

Tucker, C. M., Cadotte, M. W., Carvalho, S. B., Davies, T. J., Ferrier, S., Fritz, S. A., Grenyer, R., Helmus, M. R., Jin, L. S., Mooers, A. O., Pavoine, S., Purschke, O., Redding, D. W., Rosauer, D. F., Winter, M., & Mazel, F. (2017). A guide to phylogenetic metrics for conservation, community ecology and macroecology. *Biological Reviews*, *92*(2), 698–715. https://doi.org/10.1111/brv.12252

Urban, M. C., Travis, J. M. J., Zurell, D., Thompson, P. L., Synes, N. W., Scarpa, A., Peres-Neto, P. R., Malchow, A.-K., James, P. M. A., Gravel, D., De Meester, L., Brown, C., Bocedi, G., Albert, C. H., Gonzalez, A., & Hendry, A. P. (2021). Coding for life: Designing a platform for projecting and protecting global biodiversity. *BioScience*, *72*(1), 91–104. https://doi.org/10.1093/biosci/biab099

Vlachos, C., & Kofler, R. (2018). MimicrEE2: Genome-wide forward simulations of evolve and resequencing studies. *PLOS Computational Biology*, *14*(8), e1006413. https://doi.org/10.1371/journal.pcbi.1006413

Vostinar, A. E., & Ofria, C. (2019). Spatial structure can decrease symbiotic cooperation. *Artificial Life*, *24*(4), 229–249. https://doi.org/10.1162/artl_a_00273

Webb, C. O., Ackerly, D. D., McPeek, M. A., & Donoghue, M. J. (2002). Phylogenies and community ecology. *Annual Review of Ecology and Systematics*, *33*(1), 475–505. https://doi.org/10.1146/annurev.ecolsys.33.010802.150448